\documentstyle[emulateapj]{article}

\submitted{to appear in the Astrophysical Journal, Letters}

\lefthead{Hachisu and Kato}
\righthead{Light Curves of RS Ophiuchi Outbursts}

\begin{document}

\title{A THEORETICAL LIGHT-CURVE MODEL FOR THE 1985 OUTBURST 
OF RS OPHIUCHI}
\author{Izumi Hachisu}
\affil{Department of Earth Science and Astronomy, 
College of Arts and Sciences, University of Tokyo,
Komaba, Meguro-ku, Tokyo 153-8902, Japan; hachisu@chianti.c.u-tokyo.ac.jp}
\and
\author{Mariko Kato}
\affil{Department of Astronomy, Keio University, 
Hiyoshi, Kouhoku-ku, Yokohama 223-8521, Japan; mariko@educ.cc.keio.ac.jp}




\begin{abstract}
A theoretical light-curve model of the 1985 outburst of RS Ophiuchi 
is presented based on a thermonuclear runaway (TNR) model.  
The system consists of a very massive white dwarf (WD) with an
accretion disk (ACDK) and a red giant (RG).  
The early phase of the $V$ light curve is well 
reproduced only by the bloated WD photosphere of the TNR model 
on a $1.35 \pm 0.01 ~M_\odot$ WD 
while the later phase is dominated both 
by the irradiated ACDK and by the irradiated RG
underfilling the inner critical Roche lobe.
The UV light curve is also well reproduced by the same model
with the distance of 0.6 kpc to RS Oph.
The envelope mass at the optical peak is estimated to be
$2 \times 10^{-6} M_\odot$, indicating a rather high mass accretion 
rate of $1.2 \times 10^{-7} M_\odot$ yr$^{-1}$ between the 1967 and 
the 1985 outbursts.  About 90\% of the envelope mass is blown off
in the outburst wind while the residual 10\% 
($2 \times 10^{-7} M_\odot$) has been left and added to the helium
layer of the WD.  The net increasing rate of the WD mass is 
$1.2 \times 10^{-8} M_\odot$ yr$^{-1}$.  
Thus, RS Oph is certainly a strong candidate for a Type Ia supernova
progenitor.
\end{abstract}




%

\section{INTRODUCTION}
      RS Oph is one of the well observed recurrent 
novae and characterized by a long orbital period of 460 days
(\cite{dob94}),
a relatively short recurrence period of $\sim 10$---20 yrs, 
and a companion (M-giant) star underfilling the Roche lobe
but losing its mass by massive stellar winds (e.g., \cite{dob96}).
Historically, RS Oph underwent five outbursts, in 1898, 1933, 1958, 
1967, and 1985, with the light curves very similar each other 
(e.g., \cite{ros87}).   The latest 1985 outburst has been observed
at all wave lengths from radio to X-rays (e.g., papers in the 1985 
RS Oph conference proceedings ed. by \cite{bod87}).
\par
      Although there had been intense debates on the mechanism of 
RS Oph outbursts (e.g., \cite{liv86b}; \cite{web87}), 
various observational aspects
favor thermonuclear runaway (TNR) models on a very massive 
white dwarf (WD) (e.g., \cite{anu99} for a recent summary).  
Rapid decline rates of the light curves indicate 
a very massive WD close to the Chandrasekhar limit. 
Kato (1991) has first calculated RS Oph light curves 
for the WD masses of 1.33, 1.35, 1.36 and 1.37 $M_\odot$ and  
found that the light curve of the $1.36 M_\odot$ model 
is in better agreement with the observational light curve 
of RS Oph than the other lower mass models.  
\par
      However, structures of bloated envelopes on white dwarfs 
have been drastically changed (\cite{kat94}; \cite{kat99}) 
after the advent 
of the new opacity (e.g., \cite{igl96}).  Thus, we have been 
recalculating light curves of the recurrent novae and have
again determined the mass, composition, and distance
(\cite{hac99k} for T CrB; \cite{hkkm00}, 2000b for U Sco; 
\cite{hac00k} for V394 CrA).  Moreover, we have added new parts 
to Kato's (1991) simple model in order to follow later stages 
of the light curves, i.e., the irradiation effects of 
the red giant (RG) companion star and the accretion disk (ACDK).
As a result, much more reliable quantities are derived from 
the light curve fittings than Kato's ones. 
\par
     Type Ia supernovae (SNe Ia) are one of the most luminous 
explosive events of stars.  Recently, SNe Ia have been used 
as good distance indicators which provide a promising tool 
for determining cosmological parameters
because of their almost uniform maximum luminosities
(\cite{rie98}; \cite{per99}).
These both groups derived the maximum luminosities ($L_{\rm max}$)
of SNe Ia completely empirically from the shape of
the light curve (LCS) of nearby SNe Ia, and assumed
that the same $L_{\rm max}$--LCS relation holds for high red-shift
SNe Ia.  To be sure of any systematic biases, the physics
of SNe Ia must be understood completely.  By far, one of
the greatest problems facing SN Ia theorists is the lack of a real
progenitor (e.g., \cite{liv99} for a recent review).  
Finding a reliable progenitor is urgently required in SN Ia research.
Recurrent novae are probably the best candidate for this target
(e.g., Starrfield, Sparks, \& Truran 1985; Hachisu et al. 1999b;
Hachisu, Kato, \& Nomoto 1999a).   
\par
     In \S 2, we present our methods for calculating 
theoretical light curves 
based on the optically thick wind theory developed 
by Kato \& Hachisu (1994). 
Fitting results of the modeled light curves to the observations
are shown in \S 3.  Our results indicate that RS Oph is a strong 
candidate for Type Ia supernovae.
Discussion follows in \S 4 especially for the distance to RS Oph
because we obtain a rather short distance of 0.6 kpc.

\section{THEORETICAL LIGHT CURVES}
     In the TNR model, it has been established 
that the WD photosphere expands to a red giant size and
then gradually shrinks to the original 
size of the WD in quiescent phase 
(e.g., $\sim 0.0039 ~R_\odot$ for $M_{\rm WD}= 1.35 ~M_\odot$)
with the bolometric luminosity being kept near the Eddington limit. 
The visual luminosity reaches its maximum 
at the maximum expansion of the photosphere and then 
gradually decays to the level in quiescent phase.
Accordingly, the main emitting region of the WD photosphere 
moves from visual into soft X-rays through ultraviolet. 
Optically thick winds, which are blowing from the WD during
the outbursts, play a key role in determining the nova duration
because a large part of the envelope mass is blown in the outburst
wind.  The development of the WD photosphere can be followed 
by a unique sequence of steady-state, optically thick, wind solutions
as shown by Kato \& Hachisu (1994). 
\par
     We have calculated such sequences of $M_{\rm WD}= 1.3$, 
1.35, 1.36, 1.37, and $1.377 M_\odot$
by assuming the hydrogen content $X=0.70$, 0.50, 0.35, 
the helium content $Y=0.28$, 0.48, 0.63, respectively,
and the metallicity $Z=0.02$ (solar metallicity) and then, 
obtained the optical light curves.  Here,
we have used the updated OPAL opacity (\cite{igl96}).
The envelope mass is decreasing due to wind mass loss 
($\dot M_{\rm wind}$) and hydrogen shell burning 
($\dot M_{\rm nuc}$), i.e., 
\begin{equation}
{{d} \over {d t}} \Delta M = \dot M_{\rm acc} - 
\dot M_{\rm wind} - \dot M_{\rm nuc},
\label{dmdt_envelope_mass}
\end{equation}
where $\dot M_{\rm acc}$ is the mass accretion rate of the WD
and assumed to be 
$\dot M_{\rm acc} = 1 \times 10^{-7} M_\odot$ yr$^{-1}$ during
the outburst.
The photospheric temperature $T_{\rm ph}$, the photospheric radius
$R_{\rm ph}$, the photospheric velocity $v_{\rm ph}$, 
the wind mass loss rate $\dot M_{\rm wind}$, 
and the nuclear burning rate $\dot M_{\rm nuc}$ are 
unique functions of the envelope mass $\Delta M$.
Integrating equation (\ref{dmdt_envelope_mass}), 
we can follow the development of the envelope mass $\Delta M$ 
and obtain various physical quantities of the WD envelope.
The mass lost by the wind, $\Delta M_{\rm wind}$, and the mass
added to the helium layer of the WD, $\Delta M_{\rm He}$, are
calculated from
\begin{equation}
\Delta M_{\rm wind}= \int \dot M_{\rm wind} ~d t, 
~~~~~
\Delta M_{\rm He}= \int \dot M_{\rm nuc} ~d t. 
\label{wind_accum_mass}
\end{equation}
\par
\begin{figure}
\epsscale{.9}
\plotone{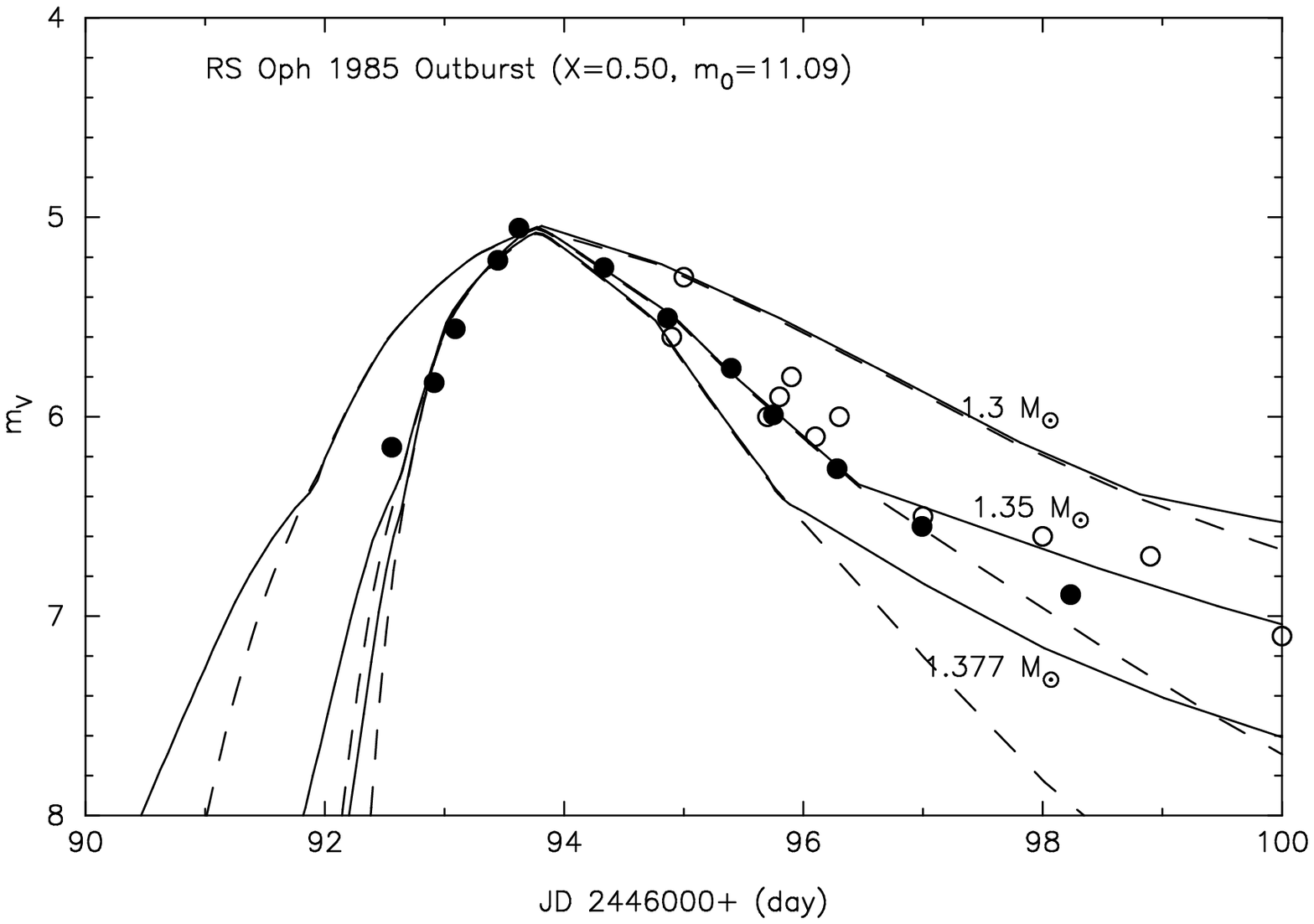}
\caption{
Model light curves are plotted against time (JD 2,446,000+) 
together with the observational points of RS Oph outbursts.
Filled circles indicate observational points with the previous
outbursts being superposed (taken from Rosino 1987). 
Open circles correspond to the observational points of the 1985 
outburst (taken from IAUC).  
Dashed lines denote the visual magnitude of the WD
photosphere only, while solid lines represent the visual magnitude
of the WD photosphere and the irradiated ACDK.
The hydrogen content of the WD envelope is assumed to be $X=0.50$.  
The apparent distance modulus is $m_0= 11.09$.
The mass of the WD is estimated to be $1.35 \pm 0.01 M_\odot$.
\label{all_mix_mass}}
\end{figure}

     Assuming a black-body photosphere,
we have calculated the $V$-magnitude of the WD photosphere 
with a response function given by Allen (1973).   
For simplicity, a limb-darkening effect is neglected.
By fitting with the observational points in the early 4 days
of the superposed $V$ light curve (e.g., \cite{ros87}) 
and the 1985 $V$ light curve,
we have determined the WD mass of $M_{\rm WD}= 1.35 \pm 0.01 M_\odot$
and the apparent distance modulus of $m_0= 11.09$ 
as shown in Figure \ref{all_mix_mass}.
The early 4 days $V$-magnitude is determined mainly by the bloated 
WD photosphere, and depends sensitively on the WD mass, 
but depends hardly on the hydrogen content, $X$, the binary parameters 
such as the mass ratio, or the accretion disk shape, so that the WD mass 
determination itself is rather robust (see also \cite{kat99}).
\placefigure{all_mix_mass}
\par
     To reproduce the $V$ light curve in the late stage
($t \sim 4$---100 days after the optical maximum), 
we have to include the contributions of the irradiated RG
and the irradiated ACDK.
As suggested by Dobrzycka et al. (1996), we have assumed 
that the RG lies well within the inner critical Roche lobe, i.e.,
\begin{equation}
R_{\rm RG} = \gamma R_2^* ~~~ (\gamma < 1),
\label{RG-size}
\end{equation}
where $R_2^*$ is the effective radius of the inner critical
Roche lobe for the RG component (e.g., \cite{egg83}) and
$\gamma$ is a numerical factor.   Here, we assume 50\% efficiency 
of the irradiation ($\eta_{\rm RG}=0.5$).
The nonirradiated photospheric temperature of the RG
is a parameter for fitting, and has been determined to be 
$T_{\rm ph, RG} = 3100$ K for $\gamma=0.4$.  
\par
     The orbit of the companion star is assumed to be circular. 
The ephemeris for the inferior conjunction of the giant is
2,444,999.9$+ 460 \times E$ (\cite{dob94}).
The light curves are calculated for six cases of the companion mass, 
i.e., $M_{\rm RG}= 0.5$, 0.6, 0.7, 0.8, 1.0, and $1.2 M_\odot$. 
Since we obtain similar light curves for all of these six masses,
we show here only the results for $M_{\rm RG}= 0.7 M_\odot$.
In this case, the separation is $a= 318.5 R_\odot$, the effective radii 
of the inner critical Roche lobes for the WD component and
the RG component are $R_1^*= 139.1 R_\odot$ 
and $R_2^*= 103.1 R_\odot$.  If $\gamma= 0.4$, then 
we have $R_2 = 0.4 R_2^* \sim 40 R_\odot$.
\par
     We have also included the luminosity coming from the ACDK
irradiated by the WD photosphere when the accretion disk
exists during the outburst.
Here, we assume 50\% efficiency of the ACDK irradiation
($\eta_{\rm DK}= 0.5$).
The viscous heating is neglected because it is rather smaller 
than that of the irradiation effects.
The temperature of the unheated surface of the ACDK including
the rim is assumed to be $T_{\rm disk}= 2000$ K.  
We have checked two other cases of $T_{\rm disk}= 1000$ and 0 K
and have found no significant differences in the light curves. 
The shape of the ACDK is assumed to be axisymmetric and 
approximated by
\begin{equation}
R_{\rm disk} = \alpha R_1^*,
\label{accretion-disk-size}
\end{equation}
and
\begin{equation}
h = \beta R_{\rm disk} \left({{\varpi} 
\over {R_{\rm disk}}} \right)^\nu,
\label{flaring-up-disk}
\end{equation}
where 
$R_{\rm disk}$ is the outer edge of the accretion disk,
$h$ the height of the surface from the equatorial plane,
$\varpi$ the distance on the equatorial plane 
from the center of the WD, 
and $\alpha$ and $\beta$ are both numerical factors which 
should be determined by fitting.  The power of $\nu$ is 
assumed to be $\nu=9/8$ from the standard disk model. 
We have checked the dependency of the light curves on the parameter 
$\nu$ by changing from $\nu=9/8$ to $\nu=2$ but cannot found
any significant differences when the disk rim is as small as
$\beta=0.01$---0.05. 
\placefigure{vmag1350va_fig2}

\section{RESULTS}
\begin{figure}
\epsscale{.9}
\plotone{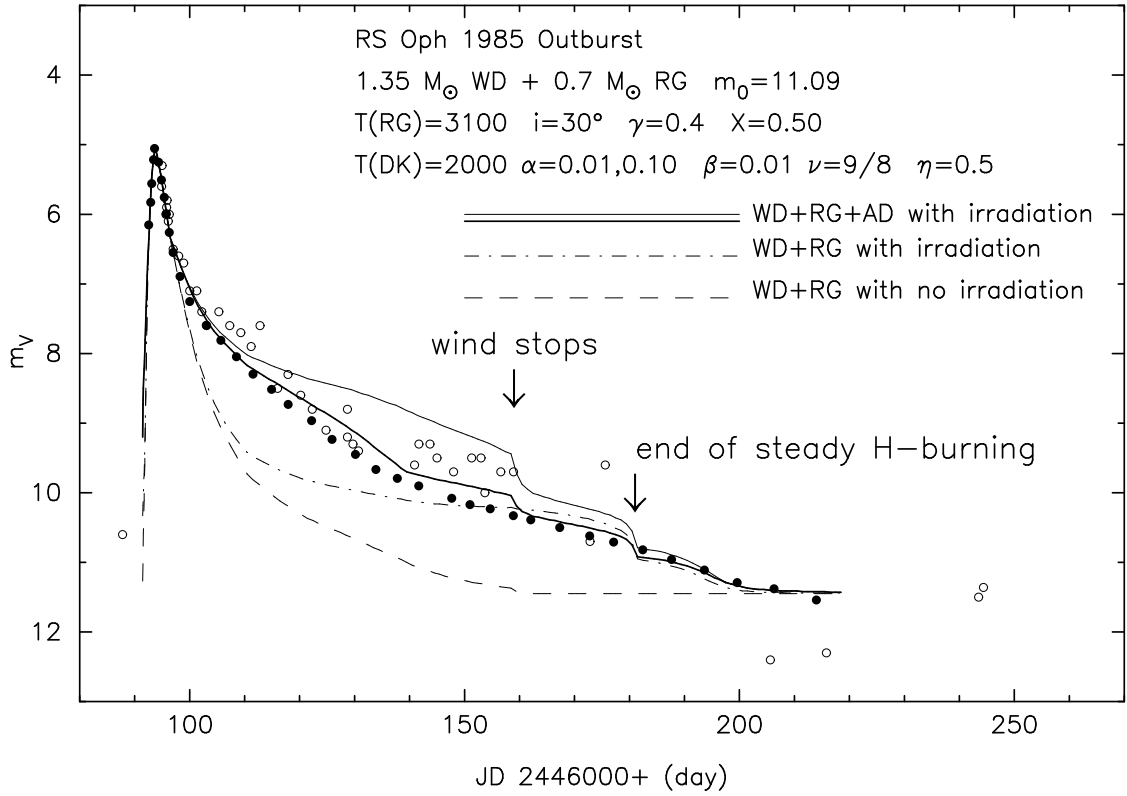}
\caption{
Same as Fig. 1 but for the entire outburst phase.
A dashed-doted line is added and it denotes the visual magnitude 
of the WD photosphere and the irradiated RG companion.
Thick and thin solid lines indicate the total visual magnitude of 
WD + RG + ACDK where the disk size is constant during the outburst,
i.e., $\alpha=0.01$ and $\beta=0.01$, and $\alpha=0.01$ and $\beta=0.01$,
respectively.  Two epochs are indicated by arrows 
with "wind stops" and "end of steady H-burning."
The optically thick wind blows during the period from the very early 
phase of the outburst (HJD 2,446,092) to $t \sim 67$ days after 
maximum (HJD 2,446,159).  The steady hydrogen shell-burning
ends at $t \sim 78$ days after maximum (HJD 2,446,180).  
\label{vmag1350va_fig2}}
\end{figure}

     The early 4 days $V$ light curve can be 
reproduced with the WD mass of $M_{\rm WD}= 1.35 \pm 0.01 M_\odot$
as already mentioned in the previous section.
The late phase $V$ light curve ($t \sim 4$---100 days after maximum) 
indicates strong irradiations both of the ACDK 
and of the RG.  If we assume a lobe-filling RG companion ($\gamma=1$),
then the irradiation of the RG is too luminous to
be compatible with the observational light curves.
On the other hand, a smaller size of the RG such as $\gamma=0.4$
($\sim 40 R_\odot$) gives a reasonable fit with the observational one 
as shown in Figure \ref{vmag1350va_fig2}.  Here, we assume
the inclination angle of $i= 30\arcdeg$ (\cite{dob94}).
\par
     To fit the late phase light curve, we have finally adopted
the disk parameters of $\alpha=0.01$ and $\beta=0.01$ as shown
in Figure \ref{vmag1350va_fig2}.  Here, the thick and thin solid 
lines denote the cases of ($\alpha=0.01$, $\beta=0.01$) and
($\alpha=0.1$, $\beta=0.01$), respectively.  Something between 
these two can roughly reproduce the light curve of the 1985 outburst 
but the case of ($\alpha=0.01$, $\beta=0.01$) is much better 
to reproduce the superposed light curve given by Rosino (1987).  
\par
     We have also calculated UV light curves with a response
function of 911\AA---3250\AA ~to fit the UV data
(\cite{sni87a}).  Figure \ref{vmag1350va_fig3}
shows three cases of UV light curves for $X=0.35$, 0.50, and 0.70,
with the parameters of $\alpha=0.01$ and $\beta=0.01$.
The calculated UV light curves can 
reproduce well the UV observations for $X=0.70$ and 
the distance to RS Oph of $d=0.57$ kpc.
Thus, we have obtained the absorption 
of $A_V= 11.09 - 5 \log(570/10) = 2.3$ for $d=0.57$ kpc.
This absorption of $A_V=2.3$ is consistent with the color excess
of $E(B-V)=0.73$ (\cite{sni87b}) because of $A_V = 3.1~E(B-V)$, 
although this distance of $\sim 0.6$ kpc is not consistent 
with another estimation of 1.6 kpc from the hydrogen column 
density (\cite{hje86}).
The UV light curves depend hardly on the disk parameters of
$\alpha$, $\beta$, or $\nu$ when $\alpha \gtrsim 0.01$ mainly
because the UV light is coming from the innermost part of
the ACDK, but depend on the irradiation efficiency of $\eta_{\rm DK}$.
We have examined that the distance becomes 0.70 kpc 
for $\eta_{\rm DK}=1.0$ or becomes 0.50 kpc for $\eta_{\rm DK}=0.25$.
\par
     The envelope mass at the optical maximum is estimated to be
$\Delta M= 2.1 \times 10^{-6} M_\odot$, which is indicating a mass 
accretion rate of $1.2 \times 10^{-7} M_\odot$ yr$^{-1}$
during the quiescent phase between the 1968 and the 1985 outbursts.
About 90\% of the envelope mass ($1.9 \times 10^{-6} M_\odot$) 
has been blown off in the optically 
thick wind and the residual 10\% ($0.2 \times 10^{-6} M_\odot$) 
has been left and added to the helium layer of the WD.  
The residual mass itself depends on both the hydrogen content $X$ and
the WD mass.  It is roughly ranging from 7\% ($X=0.70$) to 14\%($X=0.50$)
including ambiguities of the WD mass ($M_{\rm WD}=1.35\pm 0.01 M_\odot$).
Therefore, the net mass increasing rate of the WD 
is $1.2 \times 10^{-8} M_\odot$ yr$^{-1}$, which meets the condition 
for Type Ia supernova explosion 
if the WD core consists of carbon and oxygen.
Thus, we conclude that RS Oph is an immediate progenitor of
Type Ia supernova if the donor is massive enough to supply 
fuel to the WD until the WD reaches $1.378 M_\odot$ for explosion
(Nomoto,  Thielemann, \& Yokoi 1984).
\placefigure{vmag1350va_fig3}

\section{DISCUSSION}
     Assuming a tilting accretion disk around the WD, Hachisu \& Kato
(1999) have reproduced theoretically the second peak of T CrB 
outbursts.  Such a radiation induced, tilting disk instability sets in 
if the condition
\begin{eqnarray}
{{\dot M_{\rm acc}} \over {3 \times 10^{-8}~M_\odot \mbox{~yr}^{-1}}}
& \lesssim & 
\left( {{R_{\rm disk}} \over {R_\odot}} \right)^{1/2}
\left( {{L_{\rm bol}} \over {2 \times 10^{38} 
\mbox{~erg~s}^{-1}}} \right) \cr
&\times & \left( {{R_{\rm WD}} \over {0.004 ~R_\odot}} \right)^{1/2}
\left( {{M_{\rm WD}} \over {1.35 ~M_\odot}}\right)^{-1/2}
\label{radiation_condition}
\end{eqnarray}
is satisfied (\cite{sou97}).
Our estimated accretion rate of the WD in RS Oph is about 
$\dot M_{\rm acc} \sim 1.2 \times 10^{-7} M_\odot$ yr$^{-1}$, 
which does not meet the above condition, so that we do not expect 
the radiation induced instability in RS Oph system.  
Thus we can explain the reason why a second peak as seen 
in T CrB system does not appear in RS Oph system.
\par
\begin{figure}
\epsscale{.9}
\plotone{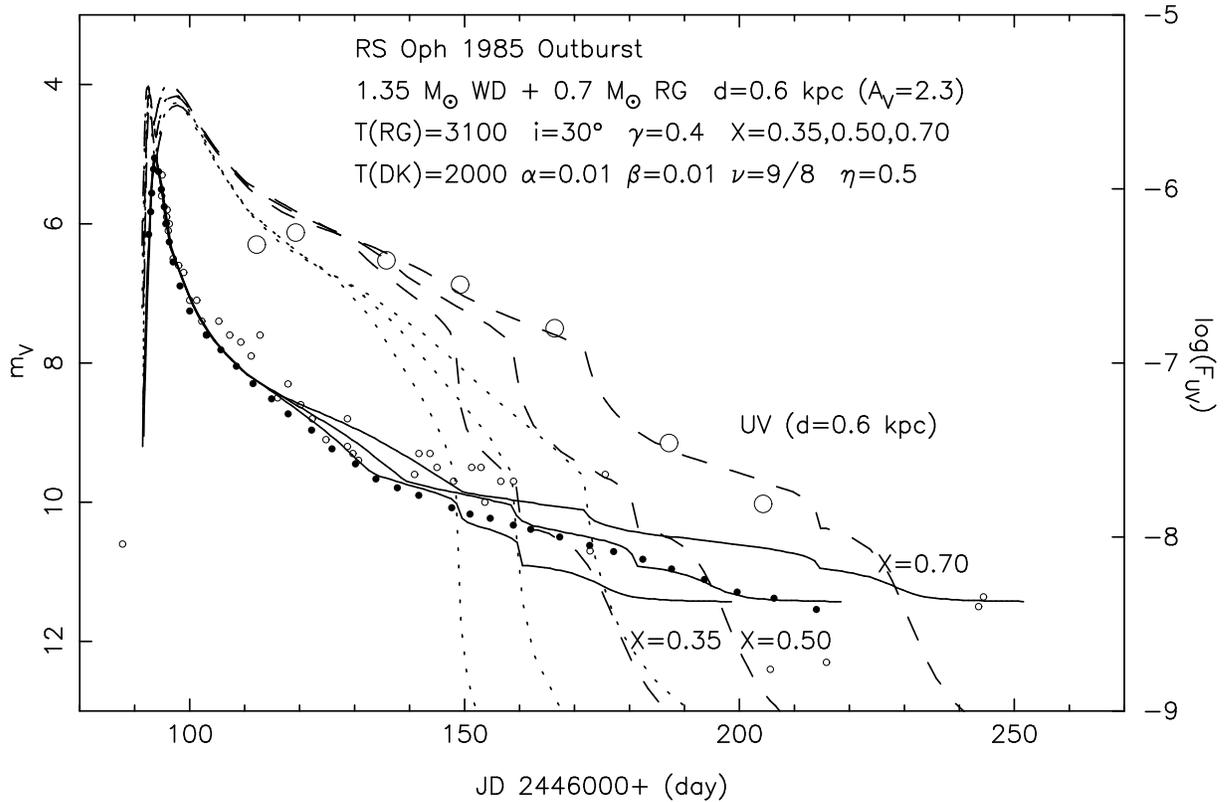}
\caption{
Same as Fig. 1 but for UV light curves.
Three modeled light curves are calculated for $X=0.35$, 0.50, and 0.70.   
Dashed lines denote total UV light from
the WD photosphere, the ACDK, and the RG companion 
but dotted lines represent UV light only from the WD photosphere.
Added large open circles are the observational points of
ultraviolet (911\AA---3250\AA) observation by Snijder (1987).
The distance to RS Oph is estimated 
to be 0.6 kpc if we assume an absorption of $A_V=2.3$.
These distance and absorption simultaneously satisfy both the $V$
light curves (solid) and the UV light curves (dashed).
\label{vmag1350va_fig3}}
\end{figure}

     Very soft X-rays were observed 251 days after the 
optical maximum in the 1985 outburst of RS Oph (\cite{mas87}).
Mason et al. estimated a black-body temperature of 
$3.5 \times 10^5$ K, a total energy flux 
of $L_{\rm X} \sim  1 \times 10^{37}$ erg~s$^{-1}~(d /1.6$ kpc)$^2$
for a hydrogen column density of $N_{\rm H}= 3 \times 10^{21}$ cm$^{-1}$,
and concluded that these soft X-rays come 
from a WD photosphere with a steady hydrogen shell-burning.
However, steady hydrogen shell-burning has stopped 
122 days after the optical maximum in our model of 
$X=0.70$ and $M_{\rm WD}= 1.35 M_\odot$ (Fig. \ref{vmag1350va_fig3}), 
thus indicating an accretion luminosity 
instead of hydrogen shell-burning.
Using the black-body temperature of $T \sim 3.5 \times 10^5$ K and
the WD radius of 0.0039 $R_\odot$, we estimate a total luminosity
of $L_{\rm X}= 4 \pi R_{\rm WD}^2 \sigma T^4 \sim 200 L_\odot$.
This low luminosity corresponds to the accretion luminosity 
of $\dot M_{\rm acc}= 2 L_{\rm X} R_{\rm WD} / G M_{\rm WD} \sim
0.5 \times 10^{-7} M_\odot$ yr$^{-1}$, which is roughly consistent with
our estimated value of 
$\dot M_{\rm acc} \sim 1.2 \times 10^{-7} M_\odot$ yr$^{-1}$.  
Then, a rather short distance of 0.45 kpc to RS Oph can be derived from 
$8 \times 10^{35} \sim  1 \times 10^{37} (d /1.6$ kpc)$^2$.
Thus, our short distance of 0.6 kpc is consistent with
the soft X-ray observation.
Moreover, the recent optical and ultraviolet observations also suggest 
100---600 $L_\odot$ for a hot component of RS Oph (e.g., \cite{dob96}).
\par
     X-rays were also observed at quiescent phase (\cite{ori93}),
but the flux is too low to be compatible with the TNR model.
One possible explanation is an absorption by the massive cool wind 
(e.g., \cite{sho96}) from the RG component as suggested by
Anupama \& Miko{\l}ajewska (1999). 
\par
     Our ejected mass ($1.9 \times 10^{-6} M_\odot$) is consistent 
with the observed ejected mass of 
(1.2---1.8)$\times 10^{-6} M_\odot$ estimated 
by Bohigas et al. (1989) except their distance of 1.6---2.0 kpc 
to RS Oph.  Using their equation (8) and $V_s=200$ and $V_e=2000$ 
km s$^{-1}$, we obtain the cool wind mass of $6 \times 10^{-5} M_\odot$,
indicating the wind mass loss rate from the RG 
of $3 \times 10^{-6} M_\odot$ yr$^{-1}$.  
It is reasonable that about one twentieth of the cool wind 
accretes to the WD (e.g., \cite{liv86a}) to power the TNR explosion. 
\par
     Hjellming et al. (1986) estimated the distance 
to RS Oph of $d= 1.6$ kpc 
from H{\sc ~I} absorption line measurements, using the hydrogen 
column density of $N_{\rm H}= (2.4 \pm 0.6) \times 10^{21}$ cm$^{-2}$ 
and the relation of $N_{\rm H}/ (T_{\rm s}~d) = 1.59 \times 10^{19}$
cm$^{-2}$ K$^{-1}$ kpc$^{-1}$ together with $T_{\rm s}=100$ K.
If a large part of this hydrogen column density stems not from 
the Galactic one but from the local one belonging to RS Oph,
the distance is overestimated.  For example, in the direction 
of the recurrent nova U Sco, Kahabka et al. (1999)
reported a hydrogen column density of (3---4)$\times 10^{21}$ cm$^{-2}$,
which is much higher than the Galactic one of $1.4 \times 10^{21}$
cm$^{-2}$.
Here, U Sco is at least 2 kpc above the Galactic plane ($b=22\arcdeg$).
If RS Oph is the same case as U Sco, the distance of 1.6 kpc is an 
upper limit and a much shorter distance is possible.

\acknowledgments
     We thank the anonymous referee for many critical comments that 
helped to improve the content of the manuscript.
This research has been supported in part by the Grant-in-Aid for
Scientific Research (09640325, 11640226) 
of the Japanese Ministry of Education, Science, Culture, and Sports.

\clearpage


%

\end{document}